\documentclass[9pt,conference]{IEEEtran}
\IEEEoverridecommandlockouts

\usepackage{cite}
\usepackage{amsmath,amssymb,amsfonts}
\usepackage{algorithmic}
\usepackage{graphicx}
\usepackage{textcomp}
\usepackage{balance}
\usepackage{booktabs}
\usepackage{hyperref}
\usepackage[table]{xcolor}
\def\BibTeX{{\rm B\kern-.05em{\sc i\kern-.025em b}\kern-.08em
    T\kern-.1667em\lower.7ex\hbox{E}\kern-.125emX}}
\makeatletter
\newcommand{\linebreakand}{%
  \end{@IEEEauthorhalign}
  \hfill\mbox{}\par
  \mbox{}\hfill\begin{@IEEEauthorhalign}
}
\makeatother
\begin{document}

\title{HDMoLE: Mixture of LoRA Experts with Hierarchical Routing and Dynamic Thresholds for Fine-Tuning LLM-based ASR Models
}
\author{
    \IEEEauthorblockN{
		\textit{Bingshen Mu}\IEEEauthorrefmark{2}\IEEEauthorrefmark{4}, 
		\textit{Kun Wei}\IEEEauthorrefmark{2}\IEEEauthorrefmark{4}, 
        \textit{Qijie Shao}\IEEEauthorrefmark{2},
        \textit{Yong Xu}\IEEEauthorrefmark{4},
        \textit{Lei Xie}\IEEEauthorrefmark{2}\IEEEauthorrefmark{1}
  }
	\IEEEauthorblockA{\IEEEauthorrefmark{2}Audio, Speech and Language Processing Group (ASLP@NPU), School of Computer Science,\\ Northwestern Polytechnical University, Xian, China\\
	\IEEEauthorblockA{\IEEEauthorrefmark{4}Tencent AI Lab, Shenzhen, China\\
 }
}\thanks{\IEEEauthorrefmark{1}: Corresponding author. This work was done when Bingshen Mu was an intern in Tencent AI lab, Shenzhen, China.}}

\maketitle

\begin{abstract}
Recent advancements in integrating Large Language Models (LLM) with automatic speech recognition (ASR) have performed remarkably in general domains.
While supervised fine-tuning (SFT) of all model parameters is often employed to adapt pre-trained LLM-based ASR models to specific domains, it imposes high computational costs and notably reduces their performance in general domains.
In this paper, we propose a novel parameter-efficient multi-domain fine-tuning method for adapting pre-trained LLM-based ASR models to multi-accent domains without catastrophic forgetting named \textit{HDMoLE}, which leverages hierarchical routing and dynamic thresholds based on combining low-rank adaptation (LoRA) with the mixture of experts (MoE) and can be generalized to any linear layer.
Hierarchical routing establishes a clear correspondence between LoRA experts and accent domains, improving cross-domain collaboration among the LoRA experts.
Unlike the static Top-K strategy for activating LoRA experts, dynamic thresholds can adaptively activate varying numbers of LoRA experts at each MoE layer.
Experiments on the multi-accent and standard Mandarin datasets demonstrate the efficacy of HDMoLE.
Applying HDMoLE to an LLM-based ASR model projector module achieves similar performance to full fine-tuning in the target multi-accent domains while using only 9.6\% of the trainable parameters required for full fine-tuning and minimal degradation in the source general domain.
\end{abstract}


\section{Introduction} \label{sec:intro}
Large language models (LLM)~\cite{ouyang2022training, achiam2023gpt, touvron2023llama, yang2024qwen2} have garnered widespread attention across various fields due to their exceptional language understanding and generation capabilities.
Extensive research has focused on exploring the potential of LLM in various fields, particularly in automatic speech recognition (ASR).
Recent developments in combining LLM with ASR have led to outstanding performance, with the paradigm of augmenting a speech foundation model with an LLM through a projector module becoming the prevailing framework for LLM-based ASR~\cite{chu2024qwen2, bai2024seed, geng2024unveiling}.
However, most LLM-based ASR models focus solely on general domain speech recognition and encounter numerous errors when confronted with speech under challenging acoustic conditions such as background noise~\cite{mu2024automatic} and speaker accents~\cite{mu2024mmger}.
While supervised fine-tuning (SFT) all parameters offers a direct approach to adapting pre-trained LLM-based ASR models to target specific acoustic domains, it requires high computational resources to retrain large LLM-based ASR models and results in considerable performance degradation in source general domains~\cite{xun24mixer, dou2024loramoe}.

Parameter-efficient fine-tuning strategies~\cite{liu2022few, neil2019parameter, xiang2021prefix, brian2021the, elad2022bitfit} seek to adapt large models to specific domains by fine-tuning only a minimal portion of model parameters, significantly reducing computational costs while minimizing the risk of catastrophic forgetting.
Low-rank adaptation (LoRA)~\cite{hu2021lora} stands out among the various strategies because it improves adaptability without changing the original model parameters.
LoRA employs low-rank decomposition to achieve weight updates via smaller matrices, allowing the model to adapt to new domains while keeping the original weight matrices unchanged, thus providing a parameter-efficient method for model adaptation.
However, more than one LoRA is needed in practical applications to fulfill user expectations.
The mixture of Experts (MoE)~\cite{shazeer2017outrageously, fedus2022review, zoph2022st, yuan2023moec, song2024u2++} is an ensemble method commonly viewed as a collection of sub-networks (experts), each focusing on different domains, with a trainable gating network (router) assigning weights to these experts.

Drawing inspiration from MoE, numerous researchers regard LoRA as a domain expert to overcome the challenges encountered by large models in real-world multi-domain scenarios.
MOELoRA~\cite{liu2023moelora} and MOA~\cite{feng2024mixture} employ domain-specific LoRA experts and explicit routing strategies to accommodate diverse domains.
SiRA~\cite{zhu2023sira} and MixLoRA~\cite{li2024mixlora} introduce sparse MoE mechanisms with specialized routing or load-balancing techniques to enhance efficiency while maintaining performance.
MoRAL~\cite{yang2024moral} tackles the challenge of adapting LLMs to new domains while enabling them to become efficient lifelong learners.
LoRAMoE~\cite{dou2024loramoe} integrates several LoRA experts through an MoE-style plugin to mitigate world knowledge forgetting in LLM during SFT.
Despite the valuable insights these studies provide into the fusion of MoE and LoRA, several challenges persist.
Firstly, the high coefficient of variation in unconstrained MoE layers reflects that the router consistently assigns larger weights to the same few experts~\cite{shazeer2017outrageously}.
The imbalance of the experts’ utilization is a typical problem in MoE, which indicates that the correspondence between LoRA experts and domains is unclear.
Secondly, the static Top-K expert selection strategy constrains the adaptability of MoE, highlighting the requirement for more dynamic expert selection strategies to address the complexities of different domains~\cite{liu2024adamole}.

This paper explores applying the mixture of LoRA experts (MoLE) to pre-trained LLM-based ASR models to improve their capabilities in handling the challenging multi-accent domains.
Accents represent deviations from standard pronunciation norms influenced by the speaker's educational background, geographical region, or native language~\cite{markl23_interspeech}, leading to significant performance degradation in pre-trained LLM-based ASR models.
To this end,  we propose a novel parameter-efficient fine-tuning method named \textit{HDMoLE} that adapts pre-trained LLM-based ASR models to multi-accent domains without catastrophic forgetting by leveraging MoLE combined with hierarchical routing and dynamic thresholds.
MoLE allows for the parameter-efficient fine-tuning of large models across multiple accent domains, effectively mitigating catastrophic forgetting.
Hierarchical routing includes global and local routing, which clarifies the correspondence between LoRA experts and accent domains while improving cross-domain collaboration among LoRA experts by assigning optimal combination weights.
Through dynamic thresholds, varying quantities of LoRA experts can be selected in each MoLE layer, with unsuitable experts discarded and higher weights reassigned to the more suitable ones.
In summary, the contributions of this paper are as follows:
\begin{itemize}
    \item To our knowledge, HDMoLE is the first attempt to explore parameter-efficient multi-domain adaptation for pre-trained LLM-based ASR models that can be applied to any linear layer.
    \item HDMoLE employs hierarchical routing and dynamic thresholds based on MoLE to clarify the correspondence between LoRA experts and accent domains, dynamically select suitable experts, and allocate higher weights to them.
    \item Extensive experiments demonstrate that HDMoLE achieves character error rate (CER) results comparable in the target multi-accent domains to full fine-tuning while using only 9.6\% of the training parameters required for full fine-tuning, with minimal degradation in the source general domain.
\end{itemize}
\begin{figure*}[!t]
  \centering
  \includegraphics[width=1.0\linewidth]{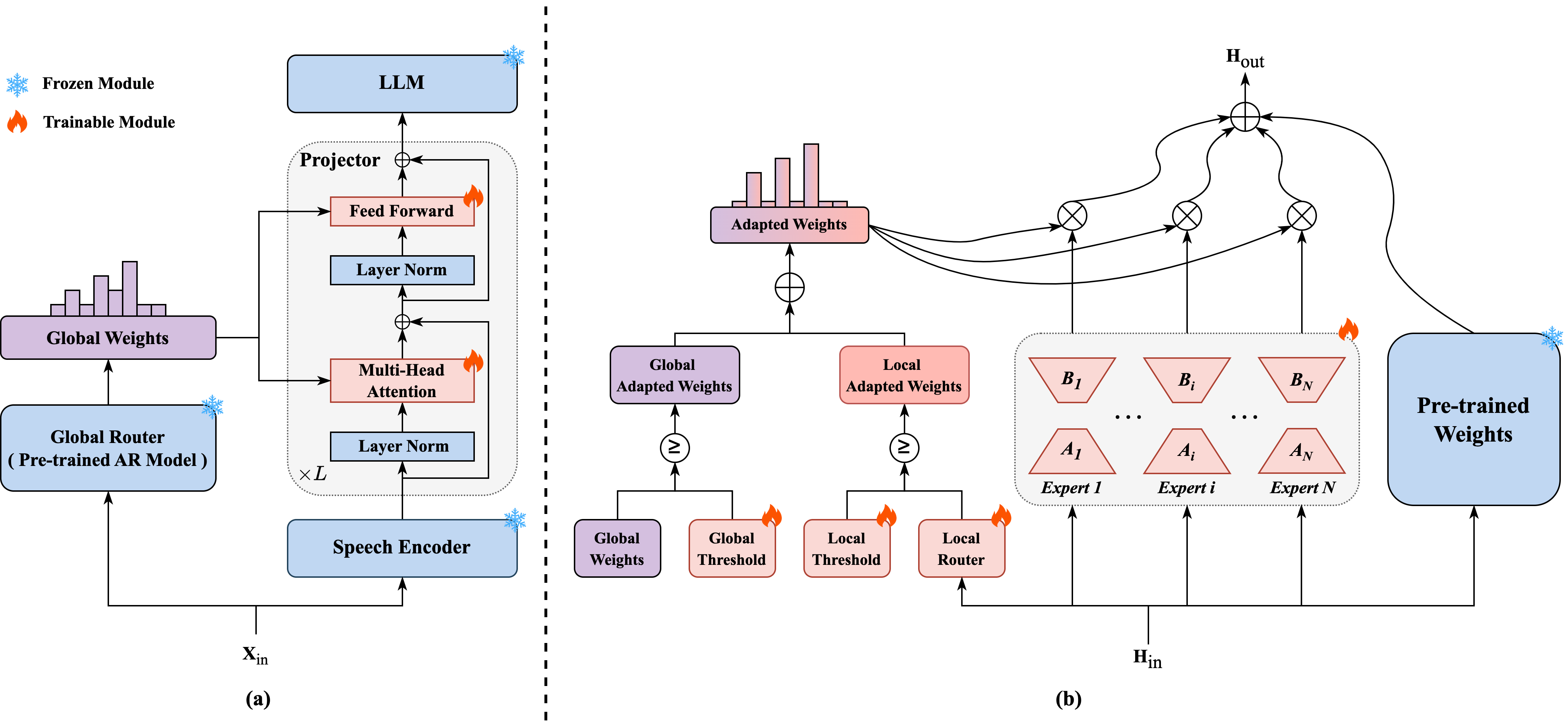}
  \caption{Overview of the proposed HDMoLE for LoRA experts combination. \textbf{(a)}: HDMoLE is integrated into the FFN and multi-head attention layers in the projector module, where the pre-trained accent recognition (AR) model provides global weights for HDMoLE; \textbf{(b)}: each HDMoLE layer employs hierarchical routing and dynamic thresholds to obtain adapted weights for combining LoRA experts.}
  \vspace{-9pt}
  \label{fig:fig1}
\end{figure*}
\section{Proposed Methods}
This section will provide a detailed introduction to HDMoLE.
While HDMoLE can be generalized to any linear layer, we use it in the projector module of a recently released LLM-based ASR model with the structure of Hubert+Baichuan2~\cite{geng2024unveiling}, chosen for its relatively compact size and alignment of accented speech and text modalities.
\subsection{Preliminaries}
\textbf{Low-Rank Adaptation.}
LoRA~\cite{hu2021lora} is an exceptional parameter-efficient fine-tuning method for adapting pre-trained large models to specific domains. 
It reduces the number of trainable parameters by updating low-rank decomposition matrix pairs while maintaining the original weights unchanged. 
Specifically, for a given linear layer with the weight matrix $\mathbf{W}_{\mathrm{0}}$, LoRA employs two low-rank matrices $\mathbf{A}$ and $\mathbf{B}$ with rank $r$, where $\mathbf{W}_{\mathrm{0}} \in \mathbb{R}^{d_{out} \times d_{in}}$, $\mathbf{A} \in \mathbb{R}^{r \times d_{in}}$, $\mathbf{B} \in \mathbb{R}^{d_{out} \times r}$, and $r \ll \min (d_{in}, d_{out})$.
With the application of LoRA, the forward process for the given linear layer $\mathbf{l}=\mathbf{W}_{\mathrm{0}}\mathbf{x} + \mathbf{b}$ can be impressed as follows:
\begin{equation}
    \mathbf{l}=\mathbf{W}_{\mathrm{0}}\mathbf{x}+\Delta \mathbf{W}\mathbf{x} + \mathbf{b} = \mathbf{W}_{\mathrm{0}}\mathbf{x}+\frac{\alpha}{r}\mathbf{BAx} + \mathbf{b},
\end{equation}
where low-rank matrices $\mathbf{A}$ and $\mathbf{B}$ are trainable while the original weight $\mathbf{W}_{\mathrm{0}}$ and bias $\mathbf{b}$ remain unchanged during training.
The matrix $\mathbf{A}$ is initialized with a random Gaussian distribution, and matrix $\mathbf{B}$ starts from zero.
Scaling $\Delta \mathbf{Wx}$ by $\alpha/r$ controls the extent of adjustments to the original weights imposed by LoRA, with $\alpha$ and $r$ representing constants.

\textbf{Mixture of Experts.}
MoE~\cite{shazeer2017outrageously, fedus2022review, zoph2022st} framework scales model capacity and complexity by incorporating multiple sub-network experts, each potentially addressing specific domains or tasks.
Within an MoE layer, $N$ independent experts $\left \{ \mathbf{E}_{i}  \right \}_{i=1}^{N}$ are coordinated by a gating network router, which applies a trainable matrix to generate a probability distribution for weighting the outputs of these experts, employing a softmax function for normalization.
For the given input vector $\mathbf{x}$, the output probability distribution $\mathbf{p}$ of the router can be impressed as:
\begin{equation}
    \mathbf{p}=\text{Softmax}(\mathbf{W}_{g}\mathbf{x}),
\end{equation}
where $\mathbf{W}_{g}$ represents the trainable weights of the gating network router.
The final output $\mathbf{y}$ from the MoE layer is a weighted sum of the outputs from the top $K$ experts:
\begin{equation}
    \mathbf{y}=\sum_{i=1}^{K}\frac{\text{TopK}(\mathbf{p}_{i})}{ {\textstyle \sum_{j=1}^{K}} \text{TopK}(\mathbf{p}_{j})}\cdot \mathbf{E}_{i}(\mathbf{x}),  
\end{equation}
where $\mathbf{p}_{i}$ and $\mathbf{p}_{j}$ are the weights of $i\text{-th}$ and $j\text{-th}$ expert in MoE layer, respectively. 
The \text{TopK} function identifies and retains the highest $K$ weights, setting the rest to zero.
The weights retained by the \text{TopK} function are normalized to ensure their sum equals one.
\subsection{Hierarchical Routing}
The original intention of MoE is that the routers determine weights for each expert according to the input sample, with higher weights given to experts focused on the input domain, thereby establishing a clear correspondence between experts and domains.
However, routers in MoE layers tend to converge to a state where it always produces large weights for early-stage well-performing experts, leading to only a handful of experts having a significant impact~\cite{xun24mixer}.
In other words, the experts assigned larger weights by the MoE router essentially remain constant for inputs across different domains.
Consequently, this imbalance of the experts' utilization problem causes ambiguity in the correspondence between experts and domains.
To address this, we propose a hierarchical routing strategy to clarify the correspondence between experts and domains.
Specifically, hierarchical routing comprises global and local routing with a pre-trained accent recognition (AR) model as the global router and individual MoE layer routers as local routers.
The global routing explicitly guides each expert to focus on a specific accent domain, clarifying the correspondence between experts and accent domains and establishing them as domain-specific experts. 
Meanwhile, the local routing implicitly guides the experts within the MoE layer to collaborate across multiple accent domains through a learnable gating network router.
The input speech features $\mathbf{X}_{\mathrm{in}}$ generate global weights $\mathbf{P}_{\mathrm{g}}$ through the global router, which can be impressed as follows:
\begin{equation}
    \mathbf{P}_\mathrm{g} = \text{Softmax}(\text{Router}_\text{Global}(\mathbf{X}_\mathrm{in})),
\end{equation}
where the global router is frozen during training and inferring.
Subsequently, the global weights are input into each MoE layer.
In each MoE layer, the hidden speech features $\mathbf{H}_\mathrm{in}$ produce local weights $\mathbf{P}_\mathrm{l}$ via the local router, which can be impressed as follows:
\begin{equation}
    \mathbf{P}_\mathrm{l} = \text{Softmax}(\text{Router}_\text{Local}(\mathbf{H}_\mathrm{in})),
\end{equation}
where the local router is a trainable linear layer.
\subsection{Dynamic Thresholds}
The standard MoE layer employs the static Top-K expert selection strategy, choosing $K$ experts with the highest weights determined by the router.
Given that each MoE layer concentrates on distinct aspects of domains, varying numbers of experts are required to participate in different MoE layers.
Therefore, we propose a dynamic threshold expert selection strategy, replacing the static Top-K approach by implementing dynamic thresholds.
The dynamic thresholds strategy allows each MoE layer to select the experts that need to be activated flexibly.
This strategy requires defining a threshold, where experts with weights exceeding the threshold are selected.
It is essential to carefully initialize the threshold, as a highly initialized threshold may cause all expert weights to fall below it, resulting in no experts being selected.
Therefore, appropriate initialization of the threshold is indispensable.
Here, the threshold is initialized at $1/N$ to ensure that at least one expert is selected.
In each HDMoLE, we assign two independent dynamic thresholds to the global and local weights, respectively.
The quantities of global and local thresholds are identical.
Through the global threshold $\tau_{g}$, we can obtain the global adapted weights $\mathbf{P}_\mathrm{ga}$, which can be impressed as follows:
\begin{equation}
    \mathbf{P}_\mathrm{ga} = \frac{\mathbb{E}(\mathbf{P}_{\mathrm{g} }\ge \tau_{g} )\cdot \mathbf{P}_{\mathrm{g} }}{ {\textstyle \sum_{i=1}^{N}} \mathbb{E}(\mathbf{P}_{\mathrm{g} }^{i}\ge \tau_{g} )\cdot \mathbf{P}_{\mathrm{g} }^{i}} \cdot \tau_{g},
\end{equation}
where $\mathbb{E}$(condition) equals one if the condition is true and zero otherwise, $\mathbf{P}_{\mathrm{g}}^{i}$ represents the weight of the $i\text{-th}$ expert in the global weights.
Additionally, scaling the adapted weights by $\tau_{g}$ guarantees that the $\tau_{g}$ remains learnable during backpropagation.
Similarly, the local threshold $\tau_{l}$ allows us to obtain the local adapted weights $\mathbf{P}_\mathrm{la}$, which can be impressed as follows:
\begin{equation}
    \mathbf{P}_\mathrm{la} = \frac{\mathbb{E}(\mathbf{P}_{\mathrm{l} }\ge \tau_{l} )\cdot \mathbf{P}_{\mathrm{l} }}{ {\textstyle \sum_{i=1}^{N}} \mathbb{E}(\mathbf{P}_{\mathrm{l} }^{i}\ge \tau_{l} )\cdot \mathbf{P}_{\mathrm{l} }^{i}} \cdot \tau_{l},
\end{equation}
where $\mathbf{P}_{\mathrm{l}}^{i}$ represents the weight of the $i\text{-th}$ expert in the local weights.
The final adapted weights $\mathbf{P}_\mathrm{a}$ are the sum of the global and local adapted weights, which can be defined as follows:
\begin{equation}
    \mathbf{P}_\mathrm{a} = \mathbf{P}_\mathrm{ga} + \mathbf{P}_\mathrm{la}.
\end{equation}
\subsection{Mixture of LoRA Experts}
MoLE substitutes conventional dense layer experts with LoRA experts, rendering it a parameter-efficient fine-tuning approach.
The final output of MoLE is a combination of the weighted outputs from the LoRA experts and the original model.
After applying hierarchical routing and dynamic thresholds, the final adapted weights $\mathbf{P}_\mathrm{a}$ are obtained.
Therefore, the output of the MoLE layer $\mathbf{H}_\mathrm{out}$ can be expressed as follows:
\begin{equation}
    \mathbf{H}_\mathrm{out} = \mathbf{W}_{0}\mathbf{H}_\mathrm{in} + \frac{\alpha}{r} \sum_{i=1}^{N} \mathbf{P}_{\mathrm{a}}^{i} \cdot \mathbf{B}_{i}\mathbf{A}_{i}\mathbf{H}_{\mathrm{in}} + \mathbf{b}.
\end{equation}
Notably, when the weight of a certain LoRA expert is simultaneously lower than the global and local thresholds, the LoRA expert's weight in the final adapted weights $\mathbf{P}_\mathrm{a}$ becomes zero.
\section{Experiments}
\begin{table*}[]
    \caption{The CER(\%) results of HDMoLE and other methods on KeSpeech (target domain) and AISHELL-2 (source domain) test datasets. The best and second-best CER are \textbf{bolded} and \underline{underlined}.}
    \label{tab:tabel1}
    \centering
\scalebox{1.0}{
\begin{tabular}{lcccc}
\toprule
\textbf{Method}     & \textbf{Finetune}                           & \textbf{Trainable Param.} & \textbf{KeSpeech CER} & \textbf{AISHELL-2 CER} \\ \midrule
Hubert+Baichuan2~\cite{geng2024unveiling}    & No                                          & -                         & 25.65                 & \textbf{3.50}                   \\
Hubert+Baichuan2~\cite{geng2024unveiling}    & Full Projector                              & 51M                       & \textbf{15.64}                 & 4.91                   \\ \midrule
LoRA~\cite{hu2021lora}                & LoRA Expert                                 & 0.31M                     & 19.98                 & 4.05                   \\
MOELoRA~\cite{liu2023moelora}             & LoRA Experts \& Local Routers               & 4.92M                     & 19.95                 & 4.66                   \\
LoRAMoE~\cite{dou2024loramoe}             & LoRA Experts \& Local Routers               & 1.99M                     & 18.76                 & 4.33                   \\
MoRAL~\cite{yang2024moral}               & LoRA Experts \& Local Routers               & 1.99M                     & 19.28                 & 4.35                   \\
MoA~\cite{feng2024mixture}                 & LoRA Experts \& Local Routers               & 4.92M                     & 18.80                 & 4.47                   \\ \midrule
\rowcolor{green!10}HDMoLE (Ours)              & LoRA Experts \& Local Routers \& Thresholds & 4.92M                     & \underline{16.58}                 & \underline{3.69}                   \\
\hspace{1em}w/o Dynamic Thresholds & LoRA Experts \& Local Routers               & 4.92M                     & 17.39                 & 3.82                   \\
\hspace{2em}w/o Local Routing      & LoRA Experts                                & 4.37M                     & 18.33                 & 3.86                   \\
\hspace{2em}w/o Global Routing     & LoRA Experts \& Local Routers               & 4.92M                     & 18.76                 & 3.94                   \\ \bottomrule
\end{tabular}
}
\vspace{-9pt}
\end{table*}
\subsection{Experimental Setup}
\textbf{Datasets.}
We conduct experiments using the multi-accent Mandarin KeSpeech~\cite{tang2021kespeech} and the standard Mandarin AISHELL-2~\cite{du2018aishell} datasets to evaluate the effectiveness of HDMoLE.
KeSpeech involves 1,542 hours of speech recorded by 27,237 speakers from 34 cities in China, and the pronunciation includes standard Mandarin and eight major accented Mandarin.
All of our HDMoLE experiments are trained solely on the KeSpeech dataset. 
We use the KeSpeech test dataset to evaluate the performance of HDMoLE in the target multi-accent domains. In contrast, the AISHELL-2 test dataset measures performance degradation in the source general domain.

\textbf{Settings.}
HDMoLE is implemented in the projector module of a pre-trained LLM-based ASR model with the Hubert+Baichuan2 structure~\cite{geng2024unveiling}.
This model uses over 11,000 hours of standard Mandarin Chinese data from four general domain corpora as training datasets, including WenetSpeech~\cite{zhang2022wenetspeech}, AISHELL-1~\cite{bu2017aishell}, AISHELL-2~\cite{du2018aishell}, and AISHELL-4~\cite{fu2021aishell}, excluding multi-accent Mandarin corpora.
The projector module is a 4-layer Transformer~\cite{vaswani2017attention} with 51M parameters, where feed-forward networks (FFN) have 2560 dimensions, multi-head self-attentions (MHSA) have 256 dimensions, and the attention head is 4.
Each HDMoLE has 8 LoRA experts, each with a rank and an alpha value of 8.
We apply HDMoLE to the FFN layers and the four weight matrices $(\mathbf{W}_\mathrm{q}, \mathbf{W}_\mathrm{k}, \mathbf{W}_\mathrm{v}, \mathbf{W}_\mathrm{o})$ of the MHSA in the projector module.
The global router pre-trained frozen AR model is a 12-layer Conformer~\cite{gulati20_interspeech} with 31M parameters trained solely on KeSpeech, where the attention head is 4, and the dimensions of FFN and MHSA are respectively set to 2048 and 256, achieving an AR accuracy of 81.44\% on the KeSpeech test dataset.
\begin{table}[]
    \caption{The CER(\%) results of HDMoLE on the KeSpeech test dataset under different AR accuracy(\%) from global routers.}
    \label{tab:tabel2}
    \centering
\begin{tabular}{l|ccc}
\toprule
\textbf{AR Accuracy}  & 62.97 & 81.44 & 100 \\ \midrule
\textbf{KeSpeech CER} & 18.11   & 16.58   & \textbf{15.35} \\ \bottomrule
\end{tabular}
\end{table}
\begin{table}[]
    \caption{The CER(\%) results of HDMoLE on the KeSpeech test dataset under different LoRA ranks.}
    \label{tab:tabel3}
    \centering
\scalebox{0.97}{
\begin{tabular}{l|ccccc}
\toprule
\textbf{LoRA Rank}        & 4     & 8     & 16    & 32     & 64     \\ \midrule
\textbf{Trainable Param.} & 2.73M & 4.92M & 9.29M & 18.05M & 35.55M \\ \midrule
\textbf{KeSpeech CER}     & 16.94 & 16.58 & 16.07 & \textbf{15.92}  & 15.98  \\ \bottomrule
\end{tabular}
}
\end{table}
\subsection{Results and Discussion}
\textbf{Main Results.}
Table~\ref{tab:tabel1} presents the CER results on the KeSpeech and AISHELL-2 test datasets across various methods.
The first line shows the inference results on the KeSpeech and AISHELL-2 test datasets using the pre-trained Hubert+Baichuan2 model, with poor performance on KeSpeech attributed to the lack of accented speech corpora during pre-training.
The second line displays the results of fully fine-tuning the Hubert+Baichuan2 model projector module. The significant improvement on the KeSpeech test dataset is the target domain topline for HDMoLE. 
Meanwhile, the regression in AISHELL-2 performance indicates that full fine-tuning causes degradation in the model’s source domain performance.
Lines 3 to 7 present the results of the original LoRA and various MoLE methods on KeSpeech and AISHELL-2, none of which match the performance of HDMoLE in both the target and source domains.
Lines 8 to 11 present the ablation study of HDMoLE, demonstrating the necessity of the two strategies employed.
In summary, HDMoLE fine-tunes the Hubert+Baichuan2 model projector module using fewer trainable parameters, achieving performance close to full fine-tuning in the target domain while minimizing regression in the source domain.

\textbf{Various Global Routers.}
Table~\ref{tab:tabel2} illustrates the influence of global routers with different performances on the efficacy of HDMoLE.
Our experiments involved three kinds of global routers: non-convergent AR model, convergent AR model, and ground-truth accent labels.
We observe that improvements in the global router’s performance lead to enhanced performance for HDMoLE.
This phenomenon indicates that improved performance of the global router clarifies the correspondence between LoRA experts and accent domains, enabling each expert to concentrate more effectively on its specific domain.
\begin{figure}[]
  \centering
  \includegraphics[width=0.9\linewidth]{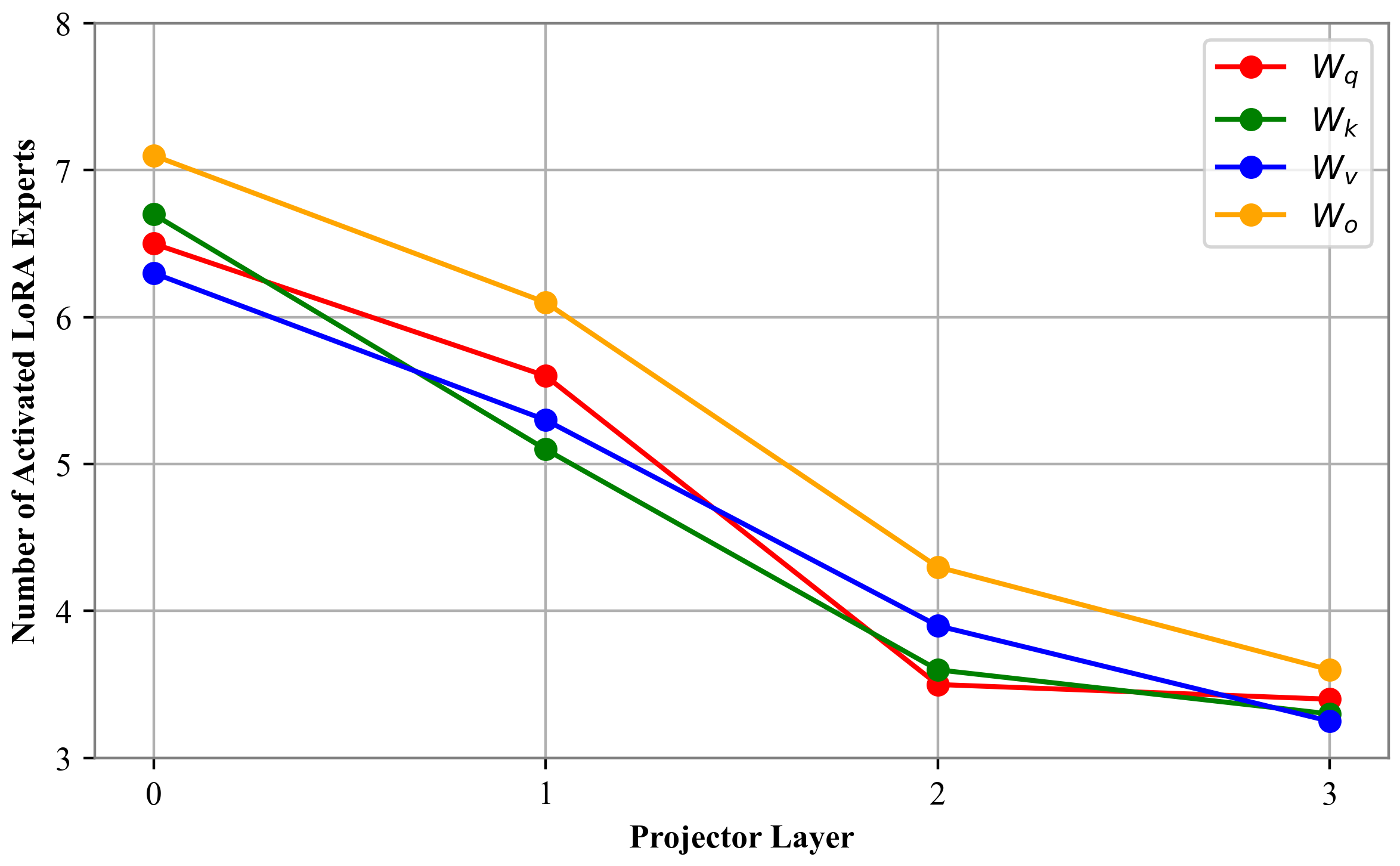}
  \caption{The number of activated LoRA experts in the MHSA four modules using HDMoLE within each projector layer. The legend refers to four modules where HDMoLE is employed.}
   \label{fig:fig2}
\end{figure}

\textbf{Rank of LoRA Experts.}
Table~\ref{tab:tabel3} examines the impact of LoRA rank variations on the performance of HDMoLE.
With higher LoRA ranks, the parameter for each LoRA expert grows, consequently elevating the number of trainable parameters in HDMoLE, leading to noticeable improvements in performance.
However, when the LoRA rank increases beyond 16, the performance improvement of HDMoLE diminishes, and even regression may occur.

\subsection{Visualization of Dynamic Thresholds}
Figure~\ref{fig:fig2} shows the average number of activated LoRA experts for every projector layer.
HDMoLE in each projector layer concentrates on distinct aspects, requiring a dynamic activation of LoRA experts based on the layer's focus.
HDMoLE in lower projector layers tends to activate more experts, signifying that these layers are essential for processing general speech features, which are more complex and diverse, thus requiring comprehensive domain knowledge.
Conversely, HDMoLE in higher projector layers tends to activate fewer experts, indicating that these layers focus on processing domain-specific speech features, demanding more specialized domain knowledge.
\section{Conclusions}
This paper proposes a novel parameter-efficient fine-tuning method for adapting pre-trained LLM-based ASR models to the target multi-accent domains without catastrophic forgetting named HDMoLE, which employs hierarchical routing and dynamic thresholds based on MoLE and can be applied to any linear layer.
Hierarchical routing clarifies the correspondence between LoRA experts and accent domains, improving cross-domain collaboration among LoRA experts, while dynamic thresholds adaptively determine the varying number of activated LoRA experts at each MoLE layer.
Extensive experiments demonstrate that HDMoLE achieves comparable results to full fine-tuning by using only 9.6\% of the training parameters required for full fine-tuning in the target multi-accent domains while minimizing regression in the source general domain.
In the future, we will explore applying HDMoLE across all LLM-based ASR models.
\clearpage
\bibliographystyle{IEEEtran}
\balance
\bibliography{refs}

\begin{thebibliography}{10}
\providecommand{\url}[1]{#1}
\csname url@samestyle\endcsname
\providecommand{\newblock}{\relax}
\providecommand{\bibinfo}[2]{#2}
\providecommand{\BIBentrySTDinterwordspacing}{\spaceskip=0pt\relax}
\providecommand{\BIBentryALTinterwordstretchfactor}{4}
\providecommand{\BIBentryALTinterwordspacing}{\spaceskip=\fontdimen2\font plus
\BIBentryALTinterwordstretchfactor\fontdimen3\font minus \fontdimen4\font\relax}
\providecommand{\BIBforeignlanguage}[2]{{%
\expandafter\ifx\csname l@#1\endcsname\relax
\typeout{** WARNING: IEEEtran.bst: No hyphenation pattern has been}%
\typeout{** loaded for the language `#1'. Using the pattern for}%
\typeout{** the default language instead.}%
\else
\language=\csname l@#1\endcsname
\fi
#2}}
\providecommand{\BIBdecl}{\relax}
\BIBdecl

\bibitem{ouyang2022training}
L.~Ouyang, J.~Wu, X.~Jiang, D.~Almeida, C.~L. Wainwright, P.~Mishkin, C.~Zhang, S.~Agarwal, K.~Slama, A.~Ray, J.~Schulman, J.~Hilton, F.~Kelton, L.~Miller, M.~Simens, A.~Askell, P.~Welinder, P.~F. Christiano, J.~Leike, and R.~Lowe, ``{Training language models to follow instructions with human feedback},'' in \emph{Proc. NeurIPS}, 2022.

\bibitem{achiam2023gpt}
J.~Achiam, S.~Adler, S.~Agarwal, L.~Ahmad, I.~Akkaya, F.~L. Aleman, D.~Almeida, J.~Altenschmidt, S.~Altman, S.~Anadkat \emph{et~al.}, ``{{GPT-4} Technical Report},'' \emph{arXiv preprint arXiv:2303.08774}, 2023.

\bibitem{touvron2023llama}
H.~Touvron, L.~Martin, K.~Stone, P.~Albert, A.~Almahairi, Y.~Babaei, N.~Bashlykov, S.~Batra, P.~Bhargava, S.~Bhosale \emph{et~al.}, ``{Llama 2: Open Foundation and Fine-Tuned Chat Models},'' \emph{arXiv preprint arXiv:2307.09288}, 2023.

\bibitem{yang2024qwen2}
A.~Yang, B.~Yang, B.~Hui, B.~Zheng, B.~Yu, C.~Zhou, C.~Li, C.~Li, D.~Liu, F.~Huang \emph{et~al.}, ``{Qwen2 Technical Report},'' \emph{arXiv preprint arXiv:2407.10671}, 2024.

\bibitem{chu2024qwen2}
Y.~Chu, J.~Xu, Q.~Yang, H.~Wei, X.~Wei, Z.~Guo, Y.~Leng, Y.~Lv, J.~He, J.~Lin \emph{et~al.}, ``{Qwen2-Audio Technical Report},'' \emph{arXiv preprint arXiv:2407.10759}, 2024.

\bibitem{bai2024seed}
Y.~Bai, J.~Chen, J.~Chen, W.~Chen, Z.~Chen, C.~Ding, L.~Dong, Q.~Dong, Y.~Du, K.~Gao \emph{et~al.}, ``{Seed-ASR: Understanding Diverse Speech and Contexts with LLM-based Speech Recognition},'' \emph{arXiv preprint arXiv:2407.04675}, 2024.

\bibitem{geng2024unveiling}
X.~Geng, T.~Xu, K.~Wei, B.~Mu, H.~Xue, H.~Wang, Y.~Li, P.~Guo, Y.~Dai, L.~Li, M.~Shao, and L.~Xie, ``{Unveiling the Potential of LLM-Based ASR on Chinese Open-Source Datasets},'' \emph{arXiv preprint arXiv:2405.02132}, 2024.

\bibitem{mu2024automatic}
B.~Mu, P.~Guo, D.~Guo, P.~Zhou, W.~Chen, and L.~Xie, ``{Automatic Channel Selection and Spatial Feature Integration for Multi-Channel Speech Recognition Across Various Array Topologies},'' in \emph{Proc. ICASSP}, 2024, pp. 11\,396--11\,400.

\bibitem{mu2024mmger}
B.~Mu, X.~Wan, N.~Zheng, H.~Zhou, and L.~Xie, ``{MMGER: Multi-Modal and Multi-Granularity Generative Error Correction With LLM for Joint Accent and Speech Recognition},'' \emph{IEEE Signal Processing Letters}, vol.~31, pp. 1940--1944, 2024.

\bibitem{xun24mixer}
X.~Wu, S.~Huang, and F.~Wei, ``Mixture of lora experts,'' in \emph{Proc. ICLR}, 2024.

\bibitem{dou2024loramoe}
S.~Dou, E.~Zhou, Y.~Liu, S.~Gao, W.~Shen, L.~Xiong, Y.~Zhou, X.~Wang, Z.~Xi, X.~Fan, S.~Pu, J.~Zhu, R.~Zheng, T.~Gui, Q.~Zhang, and X.~Huang, ``{LoRAMoE: Alleviating World Knowledge Forgetting in Large Language Models via MoE-Style Plugin},'' in \emph{Proc. ACL}, 2024, pp. 1932--1945.

\bibitem{liu2022few}
H.~Liu, D.~Tam, M.~Muqeeth, J.~Mohta, T.~Huang, M.~Bansal, and C.~Raffel, ``{Few-Shot Parameter-Efficient Fine-Tuning is Better and Cheaper than In-Context Learning},'' in \emph{Proc. NeurIPS}, 2022.

\bibitem{neil2019parameter}
N.~Houlsby, A.~Giurgiu, S.~Jastrzebski, B.~Morrone, Q.~de~Laroussilhe, A.~Gesmundo, M.~Attariyan, and S.~Gelly, ``{Parameter-Efficient Transfer Learning for {NLP}},'' in \emph{Proc. ICML}, 2019, pp. 2790--2799.

\bibitem{xiang2021prefix}
X.~L. Li and P.~Liang, ``{Prefix-Tuning: Optimizing Continuous Prompts for Generation},'' in \emph{Proc. ACL/IJCNLP}, 2021, pp. 4582--4597.

\bibitem{brian2021the}
B.~Lester, R.~Al{-}Rfou, and N.~Constant, ``{The Power of Scale for Parameter-Efficient Prompt Tuning},'' in \emph{Proc. EMNLP}, 2021, pp. 3045--3059.

\bibitem{elad2022bitfit}
E.~B. Zaken, Y.~Goldberg, and S.~Ravfogel, ``{BitFit: Simple Parameter-efficient Fine-tuning for Transformer-based Masked Language-models},'' in \emph{Proc. ACL}, 2022, pp. 1--9.

\bibitem{hu2021lora}
E.~J. Hu, Y.~Shen, P.~Wallis, Z.~Allen{-}Zhu, Y.~Li, S.~Wang, L.~Wang, and W.~Chen, ``{LoRA: Low-Rank Adaptation of Large Language Models},'' in \emph{Proc. ICLR}, 2022.

\bibitem{shazeer2017outrageously}
N.~Shazeer, A.~Mirhoseini, K.~Maziarz, A.~Davis, Q.~V. Le, G.~E. Hinton, and J.~Dean, ``{Outrageously Large Neural Networks: The Sparsely-Gated Mixture-of-Experts Layer},'' in \emph{Proc. ICLR}, 2017.

\bibitem{fedus2022review}
W.~Fedus, J.~Dean, and B.~Zoph, ``{A Review of Sparse Expert Models in Deep Learning},'' \emph{arXiv preprint arXiv:2209.01667}, 2022.

\bibitem{zoph2022st}
B.~Zoph, I.~Bello, S.~Kumar, N.~Du, Y.~Huang, J.~Dean, N.~Shazeer, and W.~Fedus, ``{ST-MoE: Designing Stable and Transferable Sparse Expert Models},'' \emph{arXiv preprint arXiv:2202.08906}, 2022.

\bibitem{yuan2023moec}
Y.~Xie, S.~Huang, T.~Chen, and F.~Wei, ``Moec: Mixture of expert clusters,'' in \emph{Proc. AAAI}, 2023, pp. 13\,807--13\,815.

\bibitem{song2024u2++}
X.~Song, D.~Wu, B.~Zhang, D.~Zhou, Z.~Peng, B.~Dang, F.~Pan, and C.~Yang, ``{U2++ MoE: Scaling 4.7 x parameters with minimal impact on RTF},'' \emph{arXiv preprint arXiv:2404.16407}, 2024.

\bibitem{liu2023moelora}
Q.~Liu, X.~Wu, X.~Zhao, Y.~Zhu, D.~Xu, F.~Tian, and Y.~Zheng, ``{MOELoRA: An MOE-based Parameter Efficient Fine-Tuning Method for Multi-task Medical Applications},'' \emph{arXiv preprint arXiv:2310.18339}, 2023.

\bibitem{feng2024mixture}
W.~Feng, C.~Hao, Y.~Zhang, Y.~Han, and H.~Wang, ``{Mixture-of-LoRAs: An Efficient Multitask Tuning Method for Large Language Models},'' in \emph{Proc. LREC-COLING}, 2024, pp. 11\,371--11\,380.

\bibitem{zhu2023sira}
Y.~Zhu, N.~Wichers, C.-C. Lin, X.~Wang, T.~Chen, L.~Shu, H.~Lu, C.~Liu, L.~Luo, J.~Chen \emph{et~al.}, ``{SiRA: Sparse Mixture of Low Rank Adaptation},'' \emph{arXiv preprint arXiv:2311.09179}, 2023.

\bibitem{li2024mixlora}
D.~Li, Y.~Ma, N.~Wang, Z.~Cheng, L.~Duan, J.~Zuo, C.~Yang, and M.~Tang, ``{MixLoRA: Enhancing Large Language Models Fine-Tuning with LoRA based Mixture of Experts},'' \emph{arXiv preprint arXiv:2404.15159}, 2024.

\bibitem{yang2024moral}
S.~Yang, M.~A. Ali, C.-L. Wang, L.~Hu, and D.~Wang, ``{MoRAL: MoE Augmented LoRA for LLMs' Lifelong Learning},'' \emph{arXiv preprint arXiv:2402.11260}, 2024.

\bibitem{liu2024adamole}
Z.~Liu and J.~Luo, ``{AdaMoLE: Fine-Tuning Large Language Models with Adaptive Mixture of Low-Rank Adaptation Experts},'' \emph{arXiv preprint arXiv:2405.00361}, 2024.

\bibitem{markl23_interspeech}
N.~Markl and C.~Lai, ``{Everyone has an accent},'' in \emph{Proc. Interspeech}, 2023, pp. 4424--4427.

\bibitem{tang2021kespeech}
Z.~Tang, D.~Wang, Y.~Xu, J.~Sun, X.~Lei, S.~Zhao, C.~Wen, X.~Tan \emph{et~al.}, ``{KeSpeech: An Open Source Speech Dataset of Mandarin and Its Eight Subdialects},'' in \emph{Proc. NeurIPS Datasets and Benchmarks Track}, 2021.

\bibitem{du2018aishell}
J.~Du, X.~Na, X.~Liu, and H.~Bu, ``{{AISHELL-2:} Transforming Mandarin {ASR} Research Into Industrial Scale},'' \emph{arXiv preprint arXiv:1808.10583}, 2018.

\bibitem{zhang2022wenetspeech}
B.~Zhang, H.~Lv, P.~Guo, Q.~Shao, C.~Yang, L.~Xie, X.~Xu, H.~Bu, X.~Chen, C.~Zeng \emph{et~al.}, ``{{WENETSPEECH:} {A} 10000+ Hours Multi-Domain Mandarin Corpus for Speech Recognition},'' in \emph{Proc. ICASSP}, 2022, pp. 6182--6186.

\bibitem{bu2017aishell}
H.~Bu, J.~Du, X.~Na, B.~Wu, and H.~Zheng, ``{{AISHELL-1:} An open-source Mandarin speech corpus and a speech recognition baseline},'' in \emph{Proc. O-COCOSDA}, 2017, pp. 1--5.

\bibitem{fu2021aishell}
Y.~Fu, L.~Cheng, S.~Lv, Y.~Jv, Y.~Kong, Z.~Chen, Y.~Hu, L.~Xie, J.~Wu, H.~Bu, X.~Xu, J.~Du, and J.~Chen, ``{AISHELL-4:} an open source dataset for speech enhancement, separation, recognition and speaker diarization in conference scenario,'' in \emph{Proc. Interspeech}, 2021, pp. 3665--3669.

\bibitem{vaswani2017attention}
A.~Vaswani, N.~Shazeer, N.~Parmar, J.~Uszkoreit, L.~Jones, A.~N. Gomez, L.~Kaiser, and I.~Polosukhin, ``{Attention is All you Need},'' in \emph{Proc. NeurIPS}, 2017, pp. 5998--6008.

\bibitem{gulati20_interspeech}
A.~Gulati, J.~Qin, C.-C. Chiu, N.~Parmar, Y.~Zhang, J.~Yu, W.~Han, S.~Wang, Z.~Zhang, Y.~Wu, and R.~Pang, ``{Conformer: Convolution-augmented Transformer for Speech Recognition},'' in \emph{Proc. Interspeech}, 2020, pp. 5036--5040.

\end{thebibliography}
\end{document}